\documentclass{elsart}

\usepackage{graphicx}
\usepackage{dcolumn}
\usepackage{bm}
\usepackage{amssymb}

\newcommand*{\bea}{\begin{eqnarray}}
\newcommand*{\eea}{\end{eqnarray}}
\newcommand*{\be}{\begin{equation}}
\newcommand*{\ee}{\end{equation}}
\newcommand*{\pd}{\partial}
\newcommand*{\pdm}{\pd_{\mu}}

\newcommand*{\pref}[1]{(\ref{#1})}

\newcommand*{\mn}{{\mu\nu}}

\newcommand*{\nn}{\nonumber}

\begin{document}

\begin{frontmatter}

\title{Linear Covariant Gauges on the Lattice}

\author[usp]{Attilio Cucchieri}\ead{attilio@ifsc.usp.br} \hskip 2mm
\author[usp,sav,graz]{Axel Maas}\ead{axel.maas@uni-graz.at} \hskip 2mm
\author[usp,ifh]{Tereza Mendes}\ead{mendes@ifsc.usp.br}

\address[usp]{Instituto de F\'\i sica de S\~ao Carlos, Universidade de S\~ao Paulo, \\
             Caixa Postal 369, 13560-970 S\~ao Carlos, SP, Brazil}

\address[sav]{Department of Complex Physical Systems, Institute of Physics, \\
              Slovak Academy of Sciences, D\'{u}bravsk\'{a} cesta 9,
              SK-845 11 Bratislava, Slovakia}

\address[graz]{Department of Theoretical Physics, Institute of Physics,\\
              Karl-Franzens University Graz, Universit\"atsplatz 5, A-8010 Graz,
              Austria}

\address[ifh]{DESY--Zeuthen, Platanenallee 6, 15738 Zeuthen, Germany}

\begin{abstract}
Linear covariant gauges, such as Feynman gauge, are very useful in perturbative calculations.
Their nonperturbative formulation is, however, highly non-trivial. In particular, it is a
challenge to define linear covariant gauges on a lattice.
We consider a class of gauges in lattice gauge theory that coincides with the
perturbative definition of linear covariant gauges in the formal continuum limit.
The corresponding gauge-fixing procedure is described and analyzed in detail,
with an application to the pure $SU(2)$ case.
In addition, results for the gluon propagator in the two-dimensional case are given.
\end{abstract}


\end{frontmatter}


\section{Introduction}

Many direct experimental tests of QCD are performed at large energies. In this regime, the
contributions obtainable in perturbation theory are dominant. Such perturbative calculations
are often most conveniently done in linear covariant gauges, such as Feynman gauge (see for
example \cite{Bohm:yx}). On the other hand, the continuation of these gauges to the
non-perturbative regime and, in particular, their definition on the lattice is not trivial.
Only for one particular representative of this class of gauges, i.e.\ the Landau gauge,
have non-perturbative studies on the lattice been thoroughly carried out. For this gauge,
progress has been made in the understanding of the infrared sector of the theory, in
particular through study of correlation functions. This was possible by a combination of
various methods, such as
Dyson-Schwinger equations \cite{vonSmekal:1997is,von Smekal:1997vx,Alkofer:2000wg,
Lerche:2002ep,Maas:2004se,Aguilar:2004sw,Alkofer:2004it,Fischer:2006ub,Alkofer:2006gz,
Fischer:2006vf,Huber:2007kc,Boucaud:2008ji,Aguilar:2008xm,Boucaud:2008ky},
renormalization-group methods \cite{Fischer:2006vf,Pawlowski:2003hq,Braun:2007bx},
stochastic quantization \cite{Zwanziger:2001kw,Zwanziger:2002ia}, lattice calculations
\cite{Bloch:2003sk,Bowman:2004jm,Cucchieri:2004sq,Bowman:2005vx,Furui:2005bu,Sternbeck:2005tk,
Sternbeck:2005re,Bogolubsky:2005wf,Furui:2006rx,Boucaud:2006if,Cucchieri:2006tf,
Ilgenfritz:2006he,Maas:2006qw,Cucchieri:2006xi,Maas:2007uv,Cucchieri:2007zm,
Kamleh:2007ud,Bogolubsky:2007bw,Bogolubsky:2007ud,Cucchieri:2007md,Sternbeck:2007ug,
Cucchieri:2007rg,Cucchieri:2008fc} and studies \cite{Dudal:2005na,Dudal:2007cw}
based on the Gribov-Zwanziger action \cite{Zwanziger:1992qr}.

In covariant gauges beyond Landau gauge, only a few results are presently available
\cite{Giusti:1996kf,Giusti:1999im,Alkofer:2003jr,Sobreiro:2005vn,Aguilar:2007nf}.
Some of these \cite{Alkofer:2003jr,Sobreiro:2005vn} suggest that the Gribov-Zwanziger
confinement mechanism \cite{Zwanziger:2001kw,Zwanziger:2002ia,Gribov:1977wm,
Zwanziger:1993dh,Alkofer:2006fu}, proposed for Coulomb and Landau gauge, may also be at work
in the complete class of linear covariant gauges. Nevertheless, a full understanding
of how non-perturbative effects, and especially confinement, manifest themselves in
linear covariant gauges is still lacking.

The aim of this work is
to provide a non-perturbative implementation of linear covariant gauges in lattice
gauge theory that goes beyond previous attempts \cite{Giusti:1996kf}.
In Section \ref{secgd} the non-perturbative definition of linear covariant gauges
will be discussed. In Section \ref{sqlc} one possible implementation on the lattice
will be introduced. The properties of the proposed algorithm will be studied in Section
\ref{spgf}, showing that this gauge definition tends to the definition of the ordinary
linear covariant gauge in the continuum limit. Finally, as an example,
the gluon propagator is presented in Sections \ref{sgpl} and \ref{sgp}.
We consider the two and three-dimensional cases and the SU(2) gauge group
in order to study quantitatively the approach to
the continuum. Our results are summarized in Section \ref{ssum}.
Technical details for the generation of configurations using the Wilson action and
for the Landau gauge fixing can be obtained from Ref.\ \cite{Cucchieri:2006tf}.
The sets of configurations (in the 2d case) employed for this work are listed separately
for each of the three considered physical volumes in Tables \ref{tlconf}-\ref{tlconfb}.

\begin{table}
\begin{center}
\begin{tabular}{|c|c|c|c|c|c|c|c|}
\hline
$V^{1/2}$ [fm] & $N$ & $\beta$ & $1/a$ [GeV] & $a$ [fm] & Sweeps & Configs \cr
\hline
3.55 & 20 & 10 & 1.11 & 0.178 & 30 & 2623 \cr
\hline
3.55 & 30 & 21.95 & 1.66 & 0.118 & 40 & 2534 \cr
\hline
3.55 & 40 & 38.6 & 2.22 & 0.0888 & 50 & 2927 \cr
\hline
3.55 & 50 & 60 & 2.77 & 0.0710 & 60 & 4654 \cr
\hline
3.55 & 60 & 86.3 & 3.32 & 0.0592 & 70 & 3563 \cr
\hline
3.55 & 70 & 117.4 & 3.88 & 0.0507 & 80 & 4275 \cr
\hline
3.55 & 80 & 153 & 4.43 & 0.0444 & 90 & 6107 \cr
\hline
3.55 & 90 & 193.7 & 4.99 & 0.0394 & 100 & 8328 \cr
\hline
3.55 & 100 & 239 & 5.54 & 0.0355 & 110 & 7195 \cr
\hline
3.55 & 120 & 344 & 6.65 & 0.0296 & 130 & 7580 \cr
\hline
3.55 & 140 & 468 & 7.76 & 0.0254 & 150 & 5227 \cr
\hline
3.55 & 160 & 611 & 8.86 & 0.0222 & 170 & 2982 \cr
\hline
3.55 & 180 & 773 & 9.97 & 0.0197 & 190 & 2204 \cr
\hline
3.55 & 200 & 954 & 11.1 & 0.0178 & 210 & 1032 \cr
\hline
3.55 & 220 & 1155 & 12.2 & 0.0162 & 230 & 959 \cr
\hline
3.55 & 242 & 1397 & 13.4 & 0.0147 & 252 & 587 \cr
\hline
3.55 & 266 & 1689 & 14.8 & 0.0133 & 276 & 210 \cr
\hline
3.55 & 294 & 2062 & 16.3 & 0.0121 & 304 & 166 \cr
\hline
3.55 & 324 & 2505 & 18.0 & 0.0110 & 334 & 87 \cr 
\hline
\end{tabular}
\end{center}
\vskip 3mm
\caption{Sets of configurations used for this work for the smallest
physical volume in the 2d case. The lattice
spacing $a$ has been evaluated using the exact infinite-volume result
for the string tension \cite{Dosch:1978jt} and the input value
$\sqrt{\sigma} = 440$ MeV. $N$ is the lattice
size in lattice units, i.e.\ $N=V^{1/2}/a$. {\em Sweeps} indicates
the number of hybrid-overrelaxation sweeps between two consecutive
thermalized configurations. {\em Configs} represents the number
of thermalized configurations generated in each case.}
\label{tlconf}
\vskip 4mm
\end{table}

\begin{table}
\begin{center}
\begin{tabular}{|c|c|c|c|c|c|c|c|}
\hline
$V^{1/2}$ [fm] & $N$ & $\beta$ & $1/a$ [GeV] & $a$ [fm] & Sweeps & Configs \cr
\hline
7.11 & 40 & 10 & 1.11 & 0.178 & 50 & 2633 \cr
\hline
7.11 & 50 & 15.35 & 1.39 & 0.142 & 60 & 2916 \cr
\hline
7.11 & 60 & 21.9 & 1.66 & 0.119 & 70 & 3973 \cr
\hline
7.11 & 70 & 29.63 & 1.94 & 0.102 & 80 & 4263 \cr
\hline
7.11 & 80 & 38.55 & 2.22 & 0.0889 & 90 & 4466 \cr
\hline
7.11 & 90 & 48.65 & 2.49 & 0.0790 & 100 & 6507 \cr
\hline
7.11 & 100 & 59.9 & 2.77 & 0.0711 & 110 & 6534 \cr
\hline
7.11 & 120 & 86.1 & 3.32 & 0.0593 & 130 & 6785 \cr
\hline
7.11 & 140 & 117 & 3.88 & 0.0508 & 150 & 5874 \cr
\hline
7.11 & 160 & 152.7 & 4.43 & 0.0444 & 170 & 3950 \cr
\hline
7.11 & 180 & 193 & 4.99 & 0.0395 & 190 & 1744 \cr
\hline
7.11 & 200 & 238.4 & 5.54 & 0.0356 & 210 & 1391 \cr
\hline
7.11 & 220 & 288 & 6.09 & 0.0323 & 230 & 605 \cr
\hline
7.11 & 242 & 349 & 6.72 & 0.0294 & 252 & 457 \cr
\hline
7.11 & 266 & 421 & 7.38 & 0.0267 & 276 & 374 \cr
\hline
7.11 & 294 & 514.6 & 8.09 & 0.0242 & 304 & 134 \cr
\hline
7.11 & 324 & 625 & 8.99 & 0.0219 & 334 & 80 \cr
\hline
\end{tabular}
\end{center}
\vskip 3mm
\caption{Same as in Table \ref{tlconf}, but for the intermediate
physical volume.}
\label{tlconfa}
\vskip 4mm
\end{table}

\begin{table}
\begin{center}
\begin{tabular}{|c|c|c|c|c|c|c|c|}
\hline
$V^{1/2}$ [fm] & $N$ & $\beta$ & $1/a$ [GeV] & $a$ [fm] & Sweeps & Configs \cr
\hline
14.2 & 80 & 10 & 1.11 & 0.178 & 90 & 4111 \cr
\hline
14.2 & 90 & 12.55 & 1.25 & 0.158 & 100 & 4736 \cr
\hline
14.2 & 100 & 15.4 & 1.39 & 0.142 & 110 & 5455 \cr
\hline
14.2 & 120 & 21.9 & 1.66 & 0.118 & 130 & 6724 \cr
\hline
14.2 & 140 & 29.7 & 1.94 & 0.101 & 150 & 6245 \cr
\hline
14.2 & 160 & 38.6 & 2.22 & 0.0888 & 170 & 4372 \cr
\hline
14.2 & 180 & 48.8 & 2.49 & 0.0789 & 190 & 2108 \cr
\hline
14.2 & 200 & 60.1 & 2.77 & 0.0710 & 210 & 1533 \cr
\hline
14.2 & 220 & 72.6 & 3.05 & 0.0647 & 230 & 582 \cr
\hline
14.2 & 242 & 87.8 & 3.35 & 0.0588 & 252 & 596 \cr
\hline
14.2 & 266 & 103 & 3.69 & 0.0534 & 276 & 465 \cr
\hline
14.2 & 294 & 129.4 & 4.08 & 0.0483 & 304 & 119 \cr
\hline
14.2 & 324 & 157 & 4.50 & 0.0438 & 334 & 107 \cr
\hline
\end{tabular}
\end{center}
\vskip 3mm
\caption{Same as in Table \ref{tlconf}, but for the largest
physical volume.}
\label{tlconfb}
\vskip 4mm
\end{table}


\section{Non-perturbative definition of linear covariant gauges}
\label{secgd}

Linear covariant gauges are defined by the average over gauge configurations satisfying
\be
\pdm A^a_\mu(x)=\Lambda^a(x) \, ,
\label{covgauge}
\ee
for arbitrary real-valued functions $\Lambda^a(x)$. [Here $A^a_\mu(x)$ is the gluon
field and $a$ is the color index, taking values $1, 2, \ldots, N_c^2-1$ for
the SU($N_c$) gauge group.] This
average is usually done by considering the partition function
\be
{\cal Z}=\int{\cal D}\Lambda\exp\left\{-\frac{1}{2\xi}\int d^dx \, \sum_a \left[
                       \Lambda^a(x) \right]^2 \right\} \, {\cal Z}(\Lambda) \, ,
\label{pidis}
\ee
i.e.\ by using a Gaussian average of width $\sqrt{\xi}$ with the gauge-fixing
parameter $\xi$.
(Note that, with the usual convention $\hbar = c = 1$, this gauge-fixing
parameter is dimensionless for any space-time dimension $d$.)
At the same time, the mean of the distribution is the (in perturbation theory unique)
Landau-gauge configuration
\be
\pdm A^a_\mu(x)=0 \, .
\label{lgauge}
\ee
Thus, in perturbation theory, the linear covariant gauge is an average over the
complete gauge orbit [as $\Lambda^a(x)$ can take any value], with Gaussian weight
around the Landau gauge.

The problem to be solved is to introduce a non-perturbative gauge prescription
and to discretize it on a lattice.
Of course, at the non-perturbative level, effects due to Gribov copies may be difficult to
single out and to quantify.
Nevertheless, in the lattice gauge-fixing procedure introduced in the next section, the
determined gauge copy is actually unique, up to ambiguities involved in defining
the (minimal) Landau gauge and to (possible) numerical artifacts.

Let us note that previous attempts to formulate this class of linear covariant gauges
\cite{Giusti:1996kf} relied on the absence of (non-trivial) zero-modes of the Faddeev-Popov
operator on the lattice in the Landau gauge. On the other hand we know that, in the
infinite-volume limit, Landau-gauge configurations belong to the so-called first Gribov
horizon \cite{Cucchieri:2006tf,Maas:2007uv,Sternbeck:2005vs,Cucchieri:2007ta}, i.e.\ such
zero-modes do exist. Thus, it is not clear whether the procedure described in \cite{Giusti:1996kf}
is well defined. Moreover, in that case one obtains on the lattice a Faddeev-Popov operator
which is different from the corresponding continuum operator.
Here we use a different approach, which is not based on the above assumption.
This gauge-fixing procedure is described in the following section and it will be called a
{\em quasi-linear covariant gauge}.


\section{Quasi-linear covariant gauges on the lattice}\label{sqlc}

Our gauge-fixing prescription is a very direct implementation of the
prescription for performing a Gaussian average around the Landau gauge.

The first step is to transform a given (thermalized) gauge-field configuration to
Landau gauge (see for example \cite{Cucchieri:1995pn,Cucchieri:1996jm,Giusti:2001xf,
Cucchieri:2003fb}). Note that, in principle, in this step it is
necessary to fix the Landau gauge completely, i.e.\ to select a configuration within
the fundamental modular region \cite{Zwanziger:1993dh,Cucchieri:1997ns}, corresponding
to the absolute minimum of the minimizing functional. However, the practical implementation
of this step is still an open problem \cite{Bogolubsky:2005wf,Bogolubsky:2007bw}.
On the other hand, it has
been conjectured \cite{Zwanziger:2003cf} that, if one determines only correlation functions
of a finite number of field operators, it is sufficient to fix the gauge to a copy inside
the first Gribov region, i.e.\ these Gribov copies should be equivalent for very large
volumes. Hence, in our quasi-linear covariant gauge, the Landau gauge configuration
is chosen inside the first Gribov region.

The second step is to obtain a configuration that satisfies Eq.\ \pref{covgauge}, for
a given $\Lambda^a(x)$, starting from the configuration fixed to Landau gauge (in
the first step). This can be obtained by recalling that, for an infinitesimal
gauge transformation
\be
g(x) \approx \underline{1} + i \phi^b(x) \tau^b \, ,
\label{ginfinit}
\ee
the gluon field $A_\mu^a(x)$ gets modified to
\be
{A'}_\mu^{a}(x) = A_\mu^a(x) + (D_\mu^{ab}\phi^b)(x) \, .
\label{aprime}
\ee
Here we indicate with $\underline{1}$ the identity matrix,
$\tau^b$ are the generators of the SU($N_c$) Lie algebra, $D_\mu^{ab}$ is
the covariant derivative defined as
\be
D_\mu^{ab} = \pdm + g_0 f^{abc} A_\mu^a(x) \, ,
\label{dcovar}
\ee
$g_0$ is the bare coupling constant and $f^{abc}$ are the structure constants of
the gauge group. Thus, if one finds $\phi^a(x)$ as a solution of the equations
\be
\pdm \left(D_\mu^{ab} \phi^b\right)(x) = \Lambda^a(x) \, ,
\label{cvgtc}
\ee
then we have
\be
\pdm {A'}_\mu^{a}(x) = \pdm \left(A_\mu^a + D_\mu^{ab}\phi^b\right)(x)
                   = \Lambda^a(x) \, ,
\label{gc}
\ee
where we used the fact that the original gluon field $A_\mu^a(x)$ satisfies the 
Landau-gauge condition $\pdm A_\mu^a(x) =0$. 
Of course, this procedure is correct only if the gauge transformation
is small. On the other hand, as we will see below, terms that are of higher order in $\phi^b(x)$
should become smaller in the continuum limit. Let us also note that Eq.\ \pref{cvgtc} above
can easily be solved, using for example a Conjugate Gradient iterative method, since the
Landau-gauge Faddeev-Popov matrix $M = - \pdm D_\mu^{ab}$ is semi-positive definite. Of course,
for a finite lattice volume $V$, the functions $\Lambda^a(x)$ should be orthogonal to the
trivial (constant) zero-modes of $- \pdm D_\mu^{ab}$. Numerically this means that the constant
mode has to be explicitly removed from $\Lambda^a(x)$ before starting the inversion of $M$.
In the infinite-volume
limit the functions $\Lambda^a(x)$ should also be orthogonal to the non-trivial
zero-modes\footnote{We did not investigate if the presence of these non-trivial zero-modes
hampers the definition of linear covariant gauges for $\xi \neq 0$.} of $M$.

Furthermore, on the lattice, the gauge transformation is applied to the
group-valued link variables $U_\mu(x)$ and for a given $\phi^a(x)$ we need to
define a group-valued gauge transformation $g(x)$. In the SU(2) case this can
easily be done by considering
\bea
g &=& \underline{1} \cos(\phi) +\frac{i \phi^a \sin \phi}{\phi}\,\sigma^a 
\label{agmap} \\
\phi &=& \sqrt{\phi^a\phi^a} \label{agmapphi} \,,
\eea
where $\sigma^a$ are the Pauli matrices and summation over color indices is
understood.\footnote{This procedure can in principle be extended to any simple
Lie group. In particular, the determination of the generators $\phi^a$ of
the gauge transformation $g$, using the prescription \pref{cvgtc}, is valid for any
gauge group. On the other hand, a simple map relating $\phi^a$ to $g$, such
as Eq.\ \pref{agmap}, is usually not available and one should probably rely on
numerical methods in order to obtain the group element $g$.}
Clearly, in the limit of small $\phi$ one finds the infinitesimal gauge
transformation \pref{ginfinit}, if we identify the Pauli matrices with the generators
$\tau^a$ of the SU(2) Lie algebra. For finite $\phi^a(x)$, the algebra-valued fields
${A'}_\mu^{a}(x)$, obtained from the link variables after this gauge transformation,
will thus not satisfy exactly the gauge condition \pref{gc}. Nevertheless the agreement
is correct up to corrections of order ${\cal O}(a)$, becoming exact in the formal
continuum limit. Indeed, besides the usual ${\cal O}(a^2)$ errors, induced by the
definition of the gluon field, Eq.\ \pref{cvgtc} is correct only for the
infinitesimal gauge transformation \pref{ginfinit}. By considering the finite
gauge transformation \pref{agmap}, one indroduces errors of order ${\cal O}(\phi^2)$.
From Eqs.\ \pref{aprime} and \pref{dcovar} one sees that $\phi$ has dimension
$1/g_0$, i.e.\ it has mass dimension $d/2 - 2$, where $d$ is the space-time
dimension. Note that, to be more consistent, one should write
the infinitesimal gauge transformation \pref{aprime} as
$g(x) \approx \underline{1} + i g_0 \phi^b(x) \tau^b$. Thus, in the
continuum limit these errors go indeed to zero. Of course, it would be
interesting to study possible improvements for Eq.\ \pref{cvgtc}, in order
to speed-up the approach to the continuum.
This also shows that our approach is not unique and other choices
for the gauge transformation $g(x)$ are possible, leading to different subleading behaviors.

Finally, in the third step, we need to average over gauge copies satisfying \pref{covgauge}.
As said above, the quantities $\Lambda^a(x)$ are randomly chosen using a Gaussian
distribution with null mean value and width $\sqrt{\xi}$. This implies that $\Lambda^a(x)$
is unbounded. On the other hand, on the lattice, using a compact formulation, the
field $A_\mu^a(x)$ is bounded and so is its derivative $\pd_\mu A_\mu^a(x)$.
Thus, $\pdm A_\mu^a(x)$ cannot really obey a Gaussian distribution and in general
it is not possible to satisfy Eq.\ \pref{covgauge} on a discrete lattice for an
arbitrary function $\Lambda^a(x)$ (see Appendix B.2 in \cite{rank}).
Nevertheless, we can still define a gauge-fixing procedure that becomes a linear
covariant gauge in the formal continuum limit. 
To this end we note that, on the lattice, the
Gaussian distribution in Eq.\ \pref{pidis} can be written as
\be
 \!\!\!\!\!\!\!\!\!\! \int{\cal D}\Lambda \exp
      \left\{ - \frac{\beta / (2 N_c)}{2\xi} \sum_{x, a} \left[\Lambda^a(x)\right]^2
                                       \right\} \,=\, \Pi_{xa}
         \int{\cal D}\Lambda^a(x) \exp \left\{ - \frac{\beta / (2 N_c)}{2\xi}
                      \left[\Lambda^a(x)\right]^2 \right\} \, .
\ee
Here, the factor $\beta / (2 N_c)$ is necessary in order to obtain the correct continuum limit.
Indeed, working in the continuum in
a generic $d$-dimensional space, Eq.\ \pref{covgauge} can be made
dimensionless by multiplying both sides by $a^2 g_0$. By recalling that in $d$
dimensions $\beta = 2 N_c / (a^{4-d} g_0^2)$ it is clear that the lattice quantity
\be
\frac{\beta / (2 N_c)}{2\xi} \sum_{x, b} \left[ a^2 g_0 \Lambda^b(x)\right]^2
\ee
becomes
\be
\frac{1}{2\xi} \frac{1}{a^{4-d} g_0^2} \int_x \frac{d^dx}{a^d}
                                         \sum_b \left[a^2 g_0 \Lambda^b(x)\right]^2
          \;  = \; \frac{1}{2\xi} \int_x d^dx \sum_b \left[\Lambda^b(x)\right]^2
\label{eq:gausslimit}
\ee
in the formal continuum limit. Thus, on the lattice, the function
$\Lambda^b(x)$ is generated from a Gaussian distribution with
width $\sigma=\sqrt{2 N_c \xi/\beta}$ at each site and for each color.
Note that in the continuum limit $\beta \to \infty$ the width of the distribution
$\sqrt{2 N_c \xi/\beta}$ shrinks to zero. Therefore, the functions $\Lambda^b(x)$
depart less from zero and the gauge fixing can be achieved using a small gauge
transformation for nearly all gauge copies entering into the average \pref{pidis}.
Note that this process is slower for larger $\xi$, since the width of the Gaussian
scales with $\sqrt{\xi}$.

\begin{figure}
\includegraphics[width=0.5\linewidth]{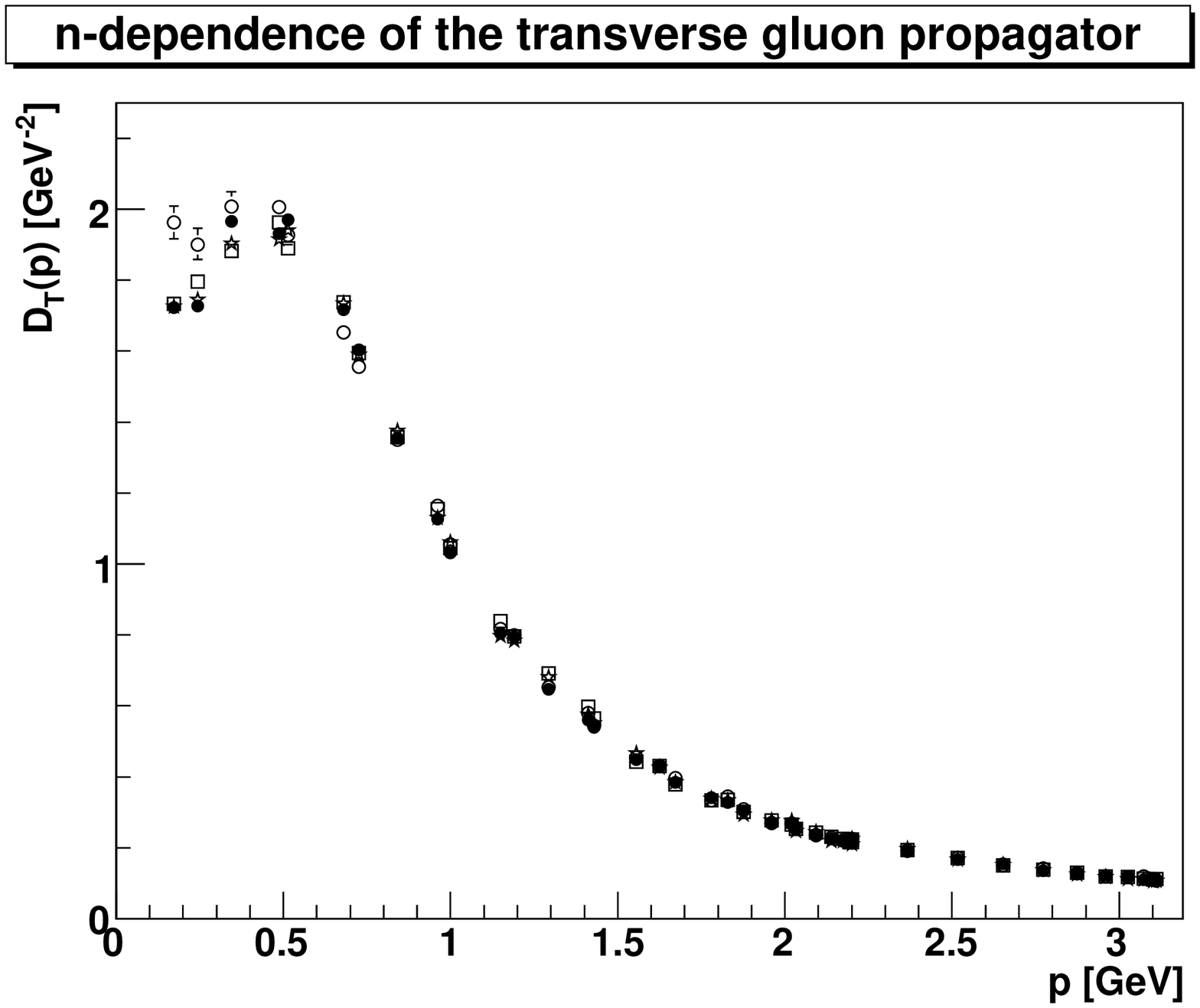}\includegraphics[width=0.5\linewidth]{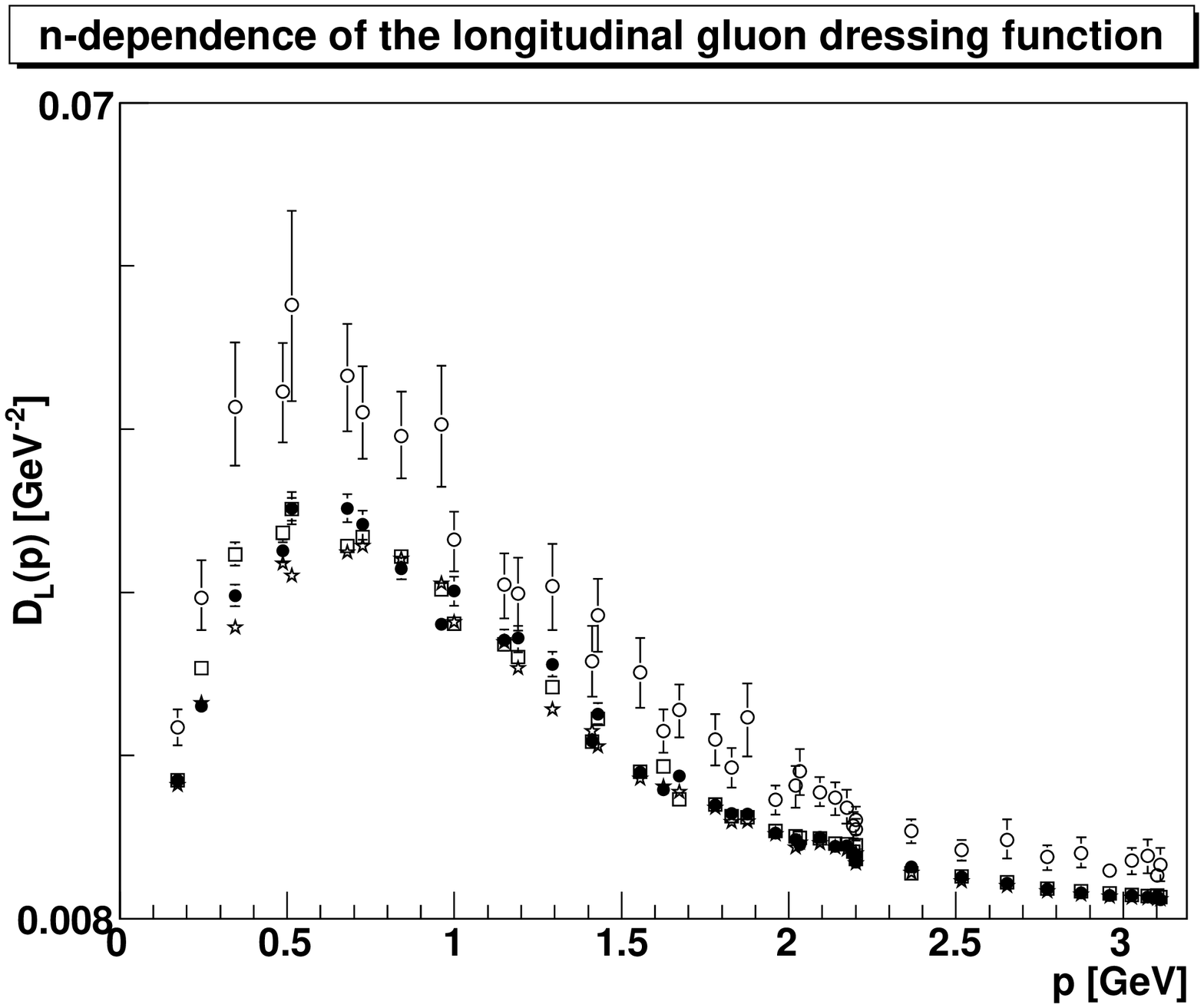}
\vskip 2mm
\caption{The transverse gluon propagator (left panel) and the longitudinal gluon
dressing function (right panel) as a function of $p$ for several values
of the number of samples $n$ on each
gauge orbit. 
Open circles correspond to $n=2$, closed circles to $n=16$, open squares
to $n=32$ and open stars to $n=1024$. Results are for a 40$^2$ lattice at $\beta=10$ for 678, 528, 480, and 228 configurations, respectively.
Note that the error bars are determined by considering each representative on a gauge
orbit as an individual measurement. For details on this evaluation of the gluon propagator
see Sections \ref{sgpl} and \ref{sgp}.}
\label{fgpn}
\vskip 4mm
\end{figure}

Finally, if one considers the partition function \pref{pidis} it is clear that, in order
to evaluate an observable in the linear covariant gauge, one
needs to perform two averages: one is the usual average over configurations ${\cal C}$,
while the other is the one along the gauge orbit ${\cal G}$. Thus, the space to be
sampled is ${\cal C}\times{\cal G}$ and errors should be estimated by regarding each
sample from ${\cal C}\times{\cal G}$ as an independent configuration. Of course, in order
to sample correctly this product space, one needs to generate enough thermalized
configurations and, for each configuration, a sufficient number $n$ of Gaussian-distributed
copies on the given gauge orbit. How large $n$ should be clearly depends on the considered
observables. For correlation functions such as the gluon propagator we find that, if the
number of thermalized configurations is large enough, then a very small number for $n$
yields a result that does not change significantly by increasing $n$, as shown in Fig.\
\ref{fgpn}. This implies that the fluctuations along a gauge orbit are smaller than
between different orbits.
 

\section{Properties of the gauge-fixing}
\label{spgf}

\begin{figure}
\includegraphics[width=0.5\linewidth]{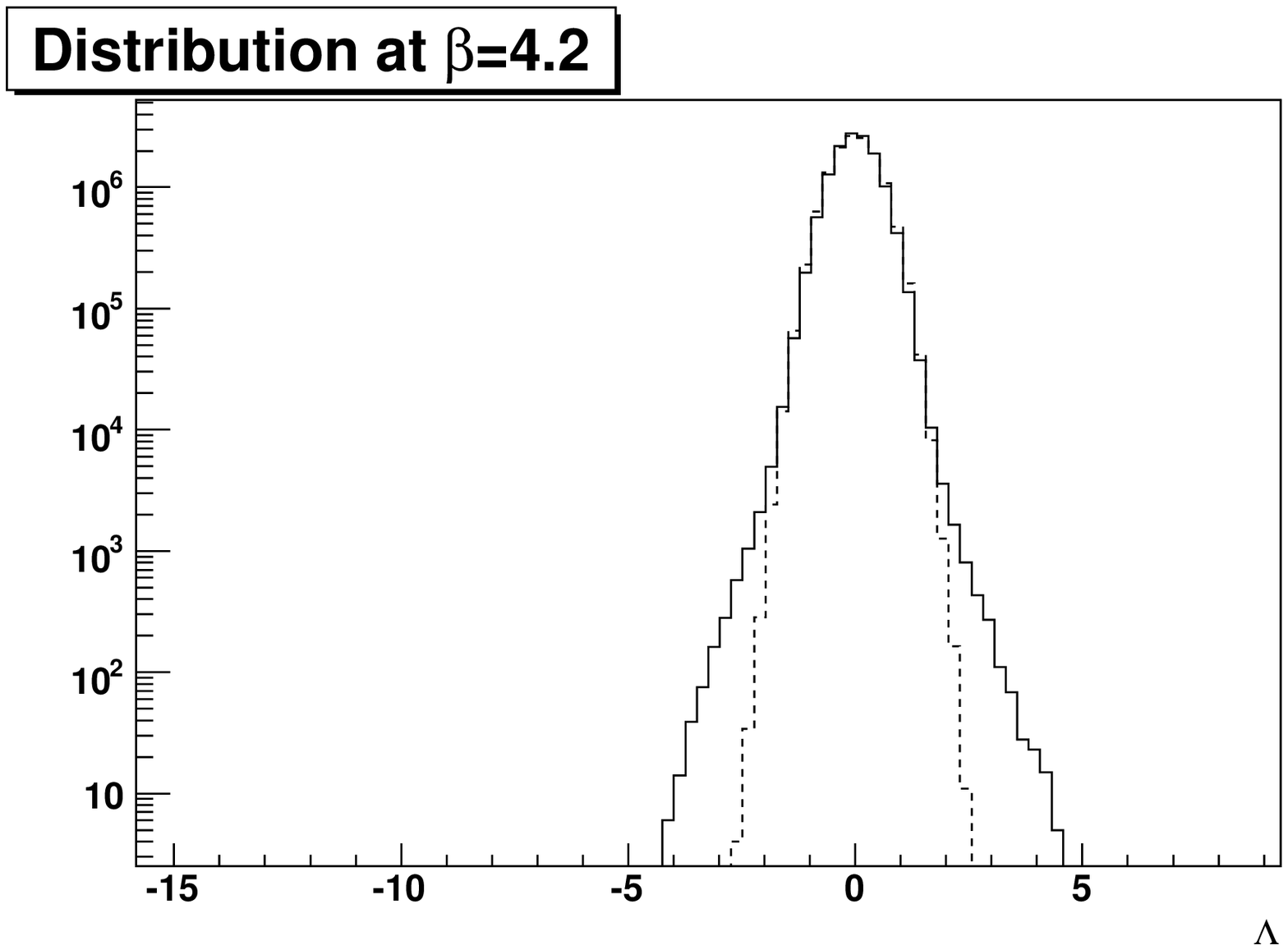}\includegraphics[width=0.5\linewidth]{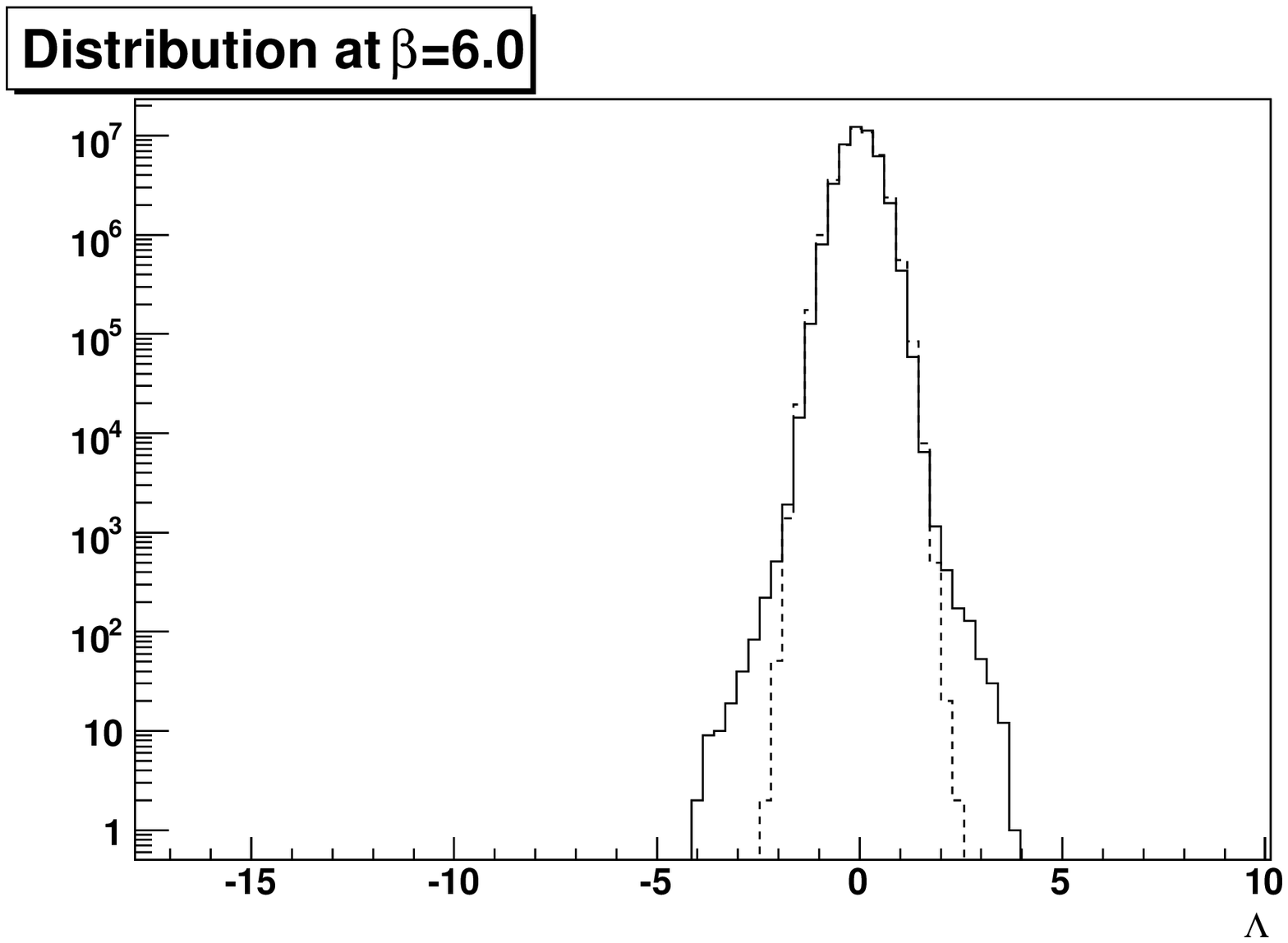}\\
\includegraphics[width=0.5\linewidth]{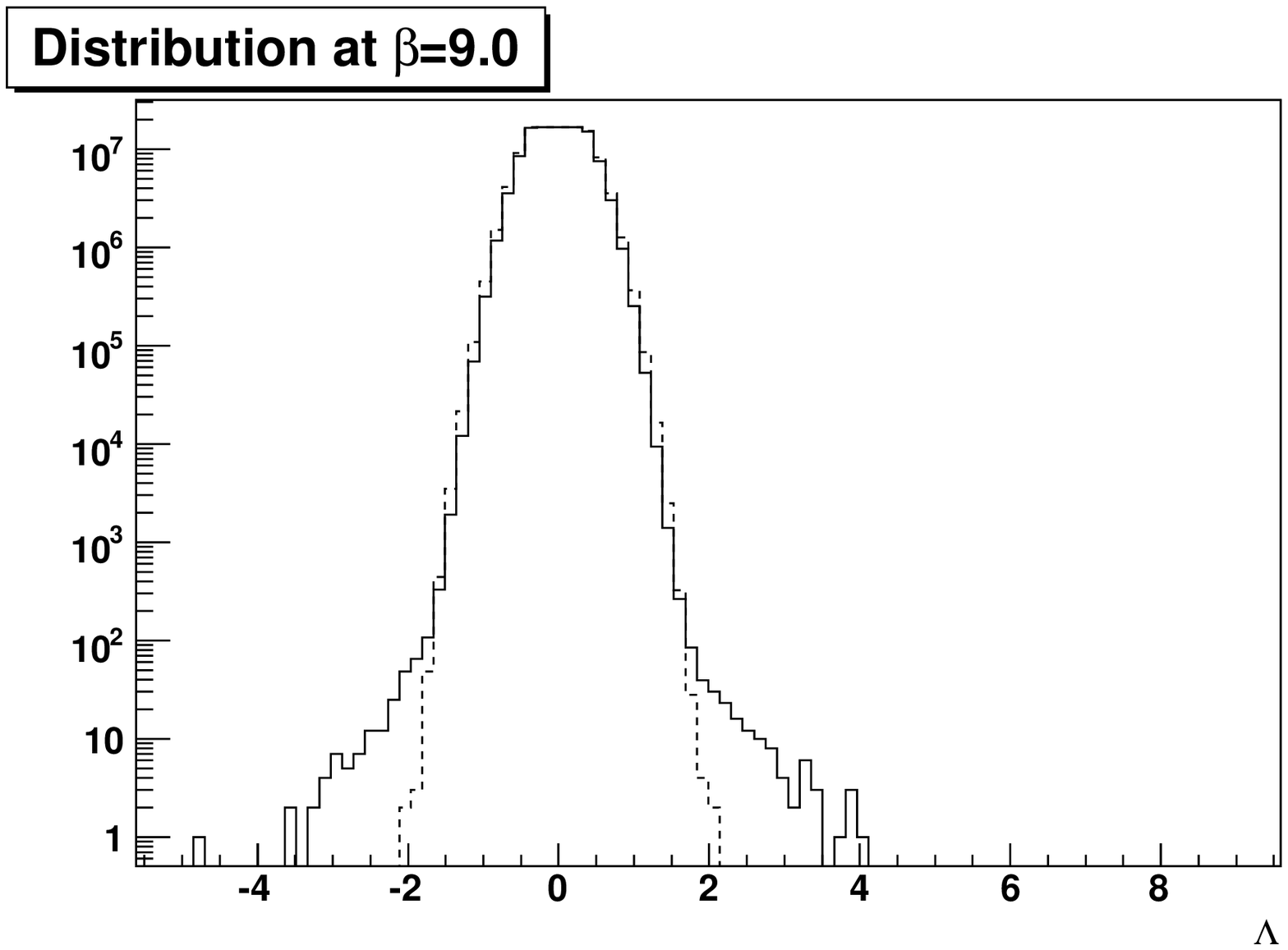}
\vskip 2mm
\caption{The measured distribution of ${\Lambda'}^{a}(x)$ (solid line), for all
lattice sites $x$, for $\beta=4.2$ and $V=12^3$
(top left panel), $\beta=6.0$ and $V=18^3$ (top right) and $\beta=9.0$ and $V=28^3$ (bottom).
In the three cases the physical volume is roughly (2.1 fm)$^3$.
In all cases we used 10 different (thermalized) configurations with $n=256$ copies
on each gauge orbit (we performed 200 thermalization sweeps for the initial
configurations and 29 sweeps between two configurations,
using the standard hybrid-overrelaxation-scheme \cite{Cucchieri:2006tf}).
The dashed line represents the initial (Gaussian) distribution with width
$\sqrt{2 N_c \xi / \beta}$. The plot range along the x-axis indicates the range of values
obtained for ${\Lambda'}^{a}(x)$.}
\label{flocal}
\vskip 4mm
\end{figure}

In this section we study the extrapolation to the continuum limit of the gauge-fixing
condition described in the previous section. The results depend, of course, 
on the value of $\xi$, i.e.\ on the width of the Gaussian distribution. Here we
consider the case $\xi=1$ (Feynman gauge) for the calculations in three dimensions.
Below, in two dimensions, we will consider the case $\xi=1/100$.

\subsection{First moment}

Our first check refers to the distributions of the one-point correlation function
${\Lambda'}^{a}(x)=\pdm {A'}_\mu^{a}(x)$, where ${A'}_\mu^{a}(x)$ is defined in
Eq.\ (\ref{aprime}). Since $\Lambda^a(x)$ is generated using a
Gaussian distribution, one should in principle obtain that the quantity ${\Lambda'}^{a}(x)$,
evaluated after fixing the configuration to the quasi-linear covariant gauge, should
also obey a Gaussian distribution with the same width. On the other hand, due to
discretization errors and to the finite-precision arithmetic employed in the
determination of $\phi^a(x)$, one can expect deviations from the original Gaussian
distribution. Thus, this check is a simple and direct way to measure the quality
of the gauge fixing and the deviation from the continuum gauge condition \pref{gc}.
Results are reported in Fig.\ \ref{flocal} and in Table \ref{tlprop} for three
different lattice volumes in 3d with (roughly) the same physical volume (2.1 fm)$^3$.
We find that in the three cases the width of the distribution of ${\Lambda'}^{a}(x)$
is in rather good agreement with the original distribution, since the measured value
$\xi_m$ of the gauge-fixing parameter is very close to 1. The deviation from 1 is actually
expected, and corresponds to the necessity of renormalizing $\xi$ in linear covariant gauges.
We also find that the gauge-fixing procedure does not introduce skewness, while in the three cases a 
non-vanishing kurtosis is clearly observable. On the other hand, this kurtosis is decreasing
with increasing $\beta$ (see Table \ref{tlprop}). Also note (see Fig.\ \ref{flocal}) that the
distribution of ${\Lambda'}^{a}(x)$ becomes visibly different from that of ${\Lambda}^{a}(x)$ for values
on the y-axis of about $10^4$ at $\beta = 4.2$, $10^3$ at $\beta = 6.0$ and $10^2$ at $\beta = 9.0$.

\begin{table}
\begin{center}
\begin{tabular}{|c|c|c|c|c|}
\hline
$\beta$ & $N^3$ & $\xi$ & $\xi_m$ & Kurtosis \cr
\hline
4.2 & $12^3$ & 1 & 0.978 & 0.539 \cr
\hline
6.0 & $18^3$ & 1 & 0.953 & 0.124 \cr
\hline
9.0 & $28^3$ & 1 & 0.951 & 0.00977 \cr
\hline
\end{tabular}
\end{center}
\vskip 3mm
\caption{Properties of the measured distribution of ${\Lambda'}^{a}(x)$.}
\label{tlprop}
\vskip 4mm
\end{table}

\begin{figure}
\includegraphics[width=0.5\linewidth]{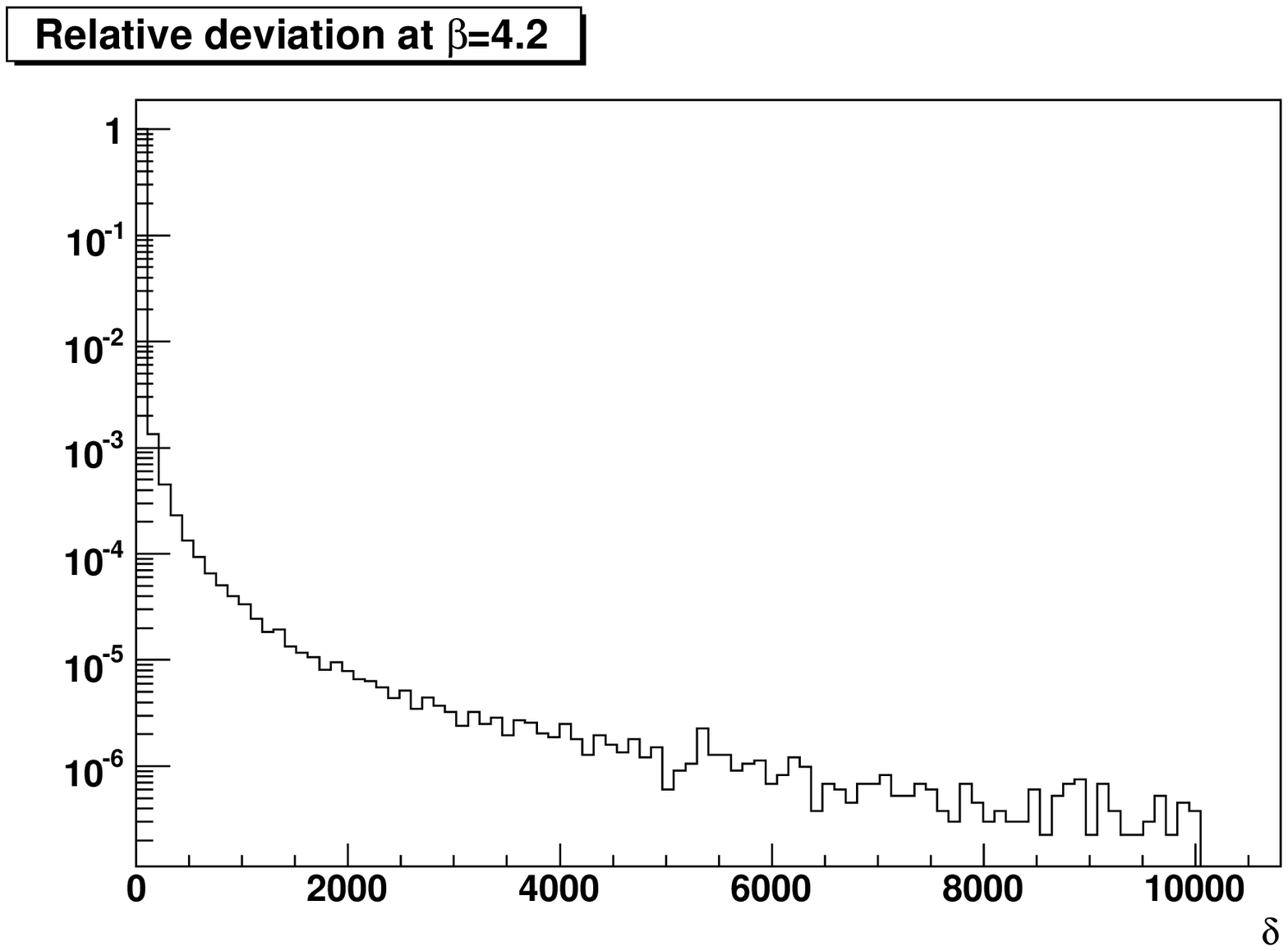}\includegraphics[width=0.5\linewidth]{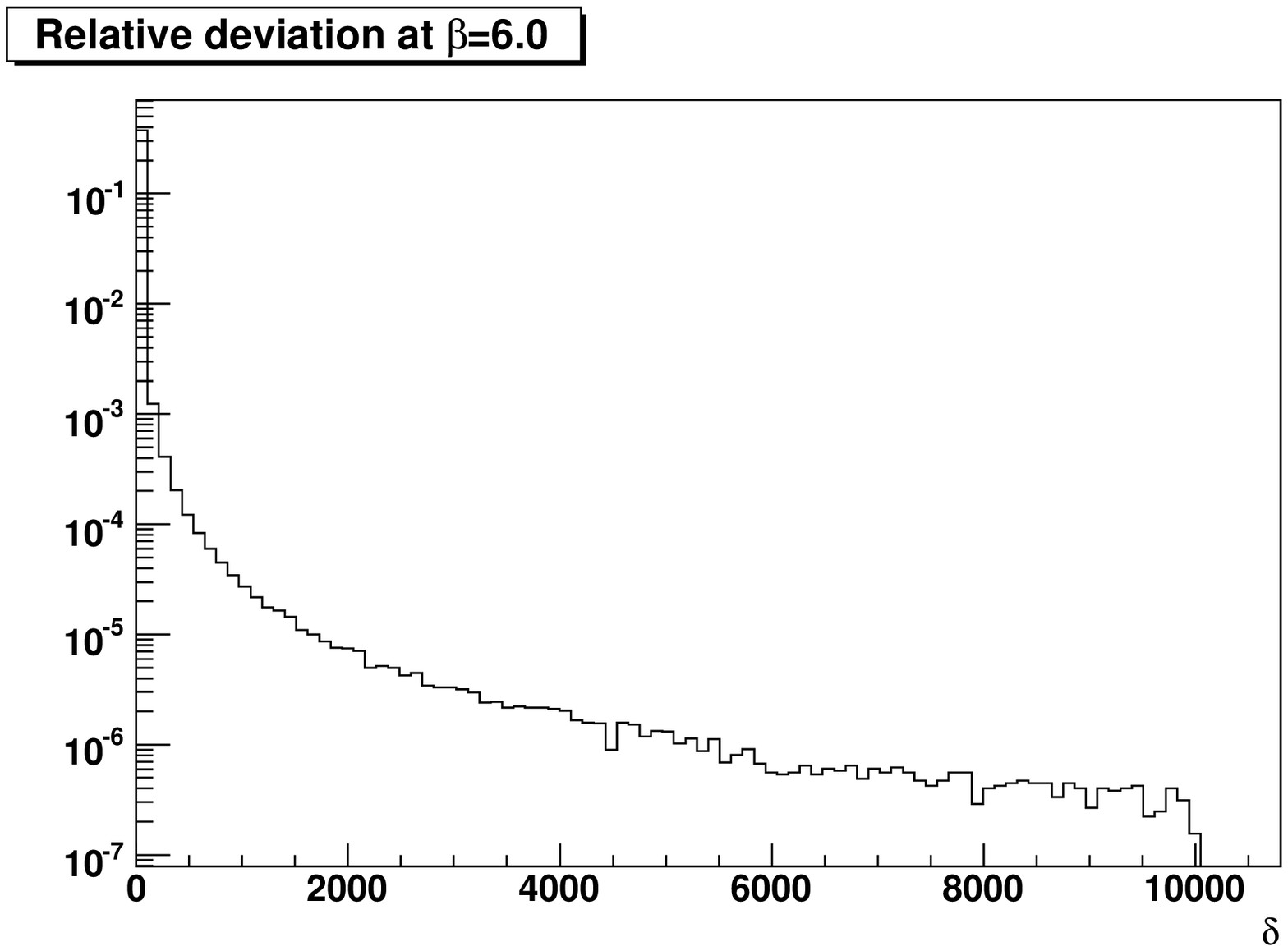}\\
\includegraphics[width=0.5\linewidth]{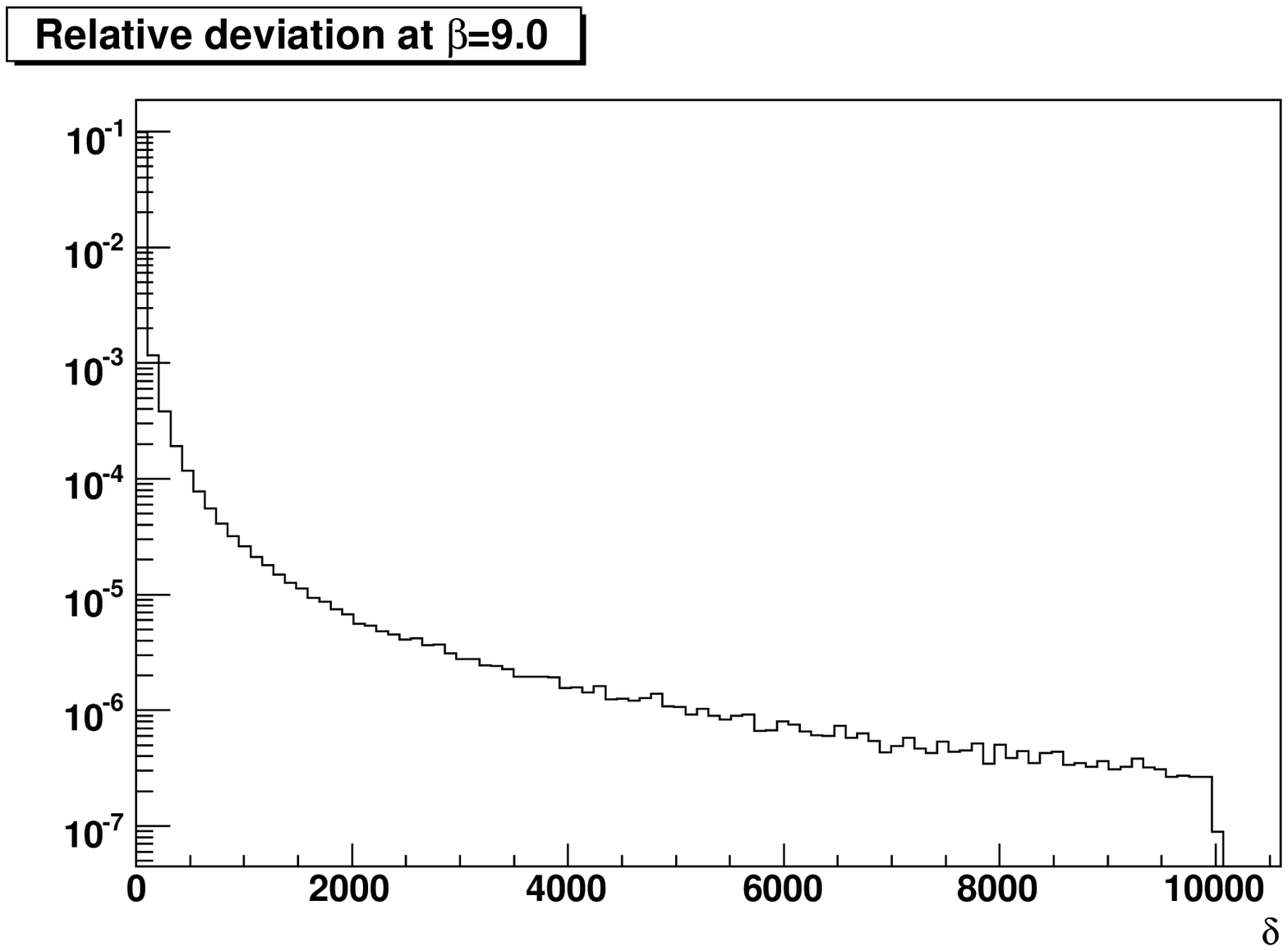}\includegraphics[width=0.46\linewidth,height=0.34\linewidth]{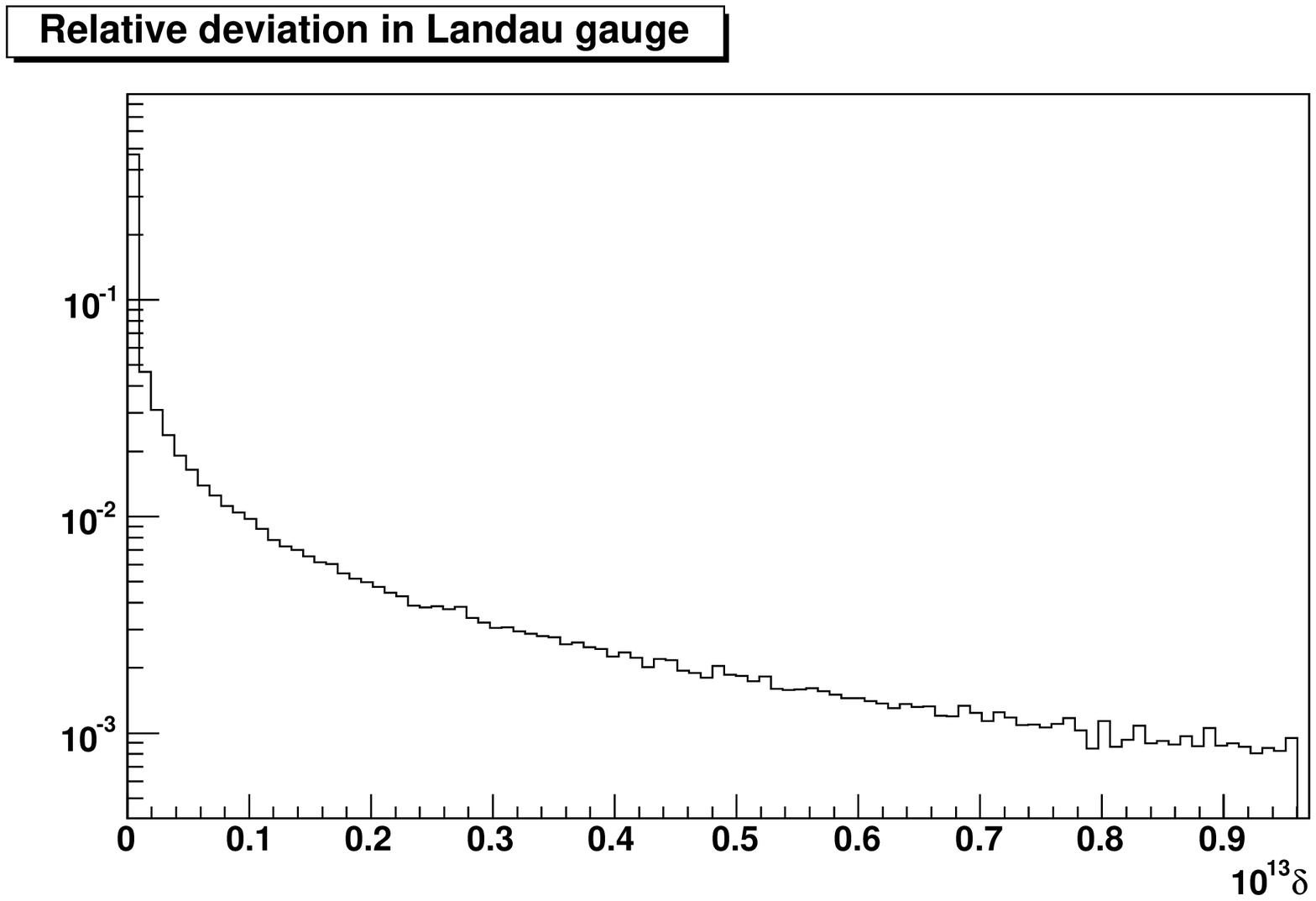}
\vskip 2mm
\caption{Histogram of the relative pointwise deviation $\delta$ for the same sets of configurations
considered in Fig.\ \ref{flocal}. Results are shown for $\beta=4.2$ in the top left panel,
for $\beta=6.0$ in the top right panel and for $\beta=9.0$ in the bottom left panel. The x-axis has
been truncated at $\delta=10^4$. For comparison, the situation in Landau gauge according to \pref{landev} is shown. The result is for a 12$^3$ lattice at $\beta=4.2$, and $\delta$ has been truncated at $10^{-13}$ with the largest $\delta$ at around 10$^{-12}$. All distributions have been normalized to a unit area.}
\label{fdev}
\vskip 4mm
\end{figure}

We also evaluated the relative (pointwise) deviation
\be
\delta=\frac{|\Lambda^a(x)-\pdm {A'}_\mu^a(x)|}{|\Lambda^a(x)|} \, .
\ee
Results are shown in Fig.\ \ref{fdev} for the same set of configurations
considered in Fig.\ \ref{flocal}. Clearly we always find sites where $\delta$
is very large. However, the maximal value of $\delta$ as well as
the number of sites with large relative deviation, both
in absolute number and normalized to the total number of sites, is decreasing with
increasing $\beta$, even though this is not clearly visible from the plots. E.\ g.,
at the cutoff value $\delta=10^4$ considered in Fig.\ \ref{fdev},
the value in the y-axis is about $6\times 10^{-6}$, $4\times 10^{-6}$ and $3\times 10^{-6}$,
respectively at $\beta=4.2$, 6.0 and 9.0. Since the histograms are normalized to a
unit area, this implies that the probability of such a large deviation is reduced
at large $\beta$.
A possible source for the very large deviations is likely related to functions $\Lambda^a$
that exceed the range that can be covered by $A_\mu^a$ and $\pd_\mu A_\mu^a$ using
the compact formulation on the lattice.

We have also checked that in Landau gauge the distribution for 
\be
\delta=|\pdm {A}_\mu^a(x)|\label{landev}
\ee
is quite similar to those shown in Fig.\ \ref{fdev}, although the absolute size of $\delta$
is then much smaller and of the order of the quality of the gauge-fixing. This is also displayed in figure \ref{fdev}.

\subsection{Second moment: The longitudinal gluon propagator}
\label{sgpl}

We have seen in the above section that,
if one considers single-site properties, the quasi-linear covariant gauge indeed
seems to approach the usual linear covariant gauge in the limit $\beta\to\infty$. On the
other hand, we should check if the same is true for all correlation functions of
$\pdm A_\mu^a(x)$. Here we consider only the two-point correlation functions of $\pd_\mu
A_\mu^a(x)$, since higher correlation functions are statistically very noisy \cite{Cucchieri:2006tf}.

\begin{figure*}
\includegraphics[width=\linewidth]{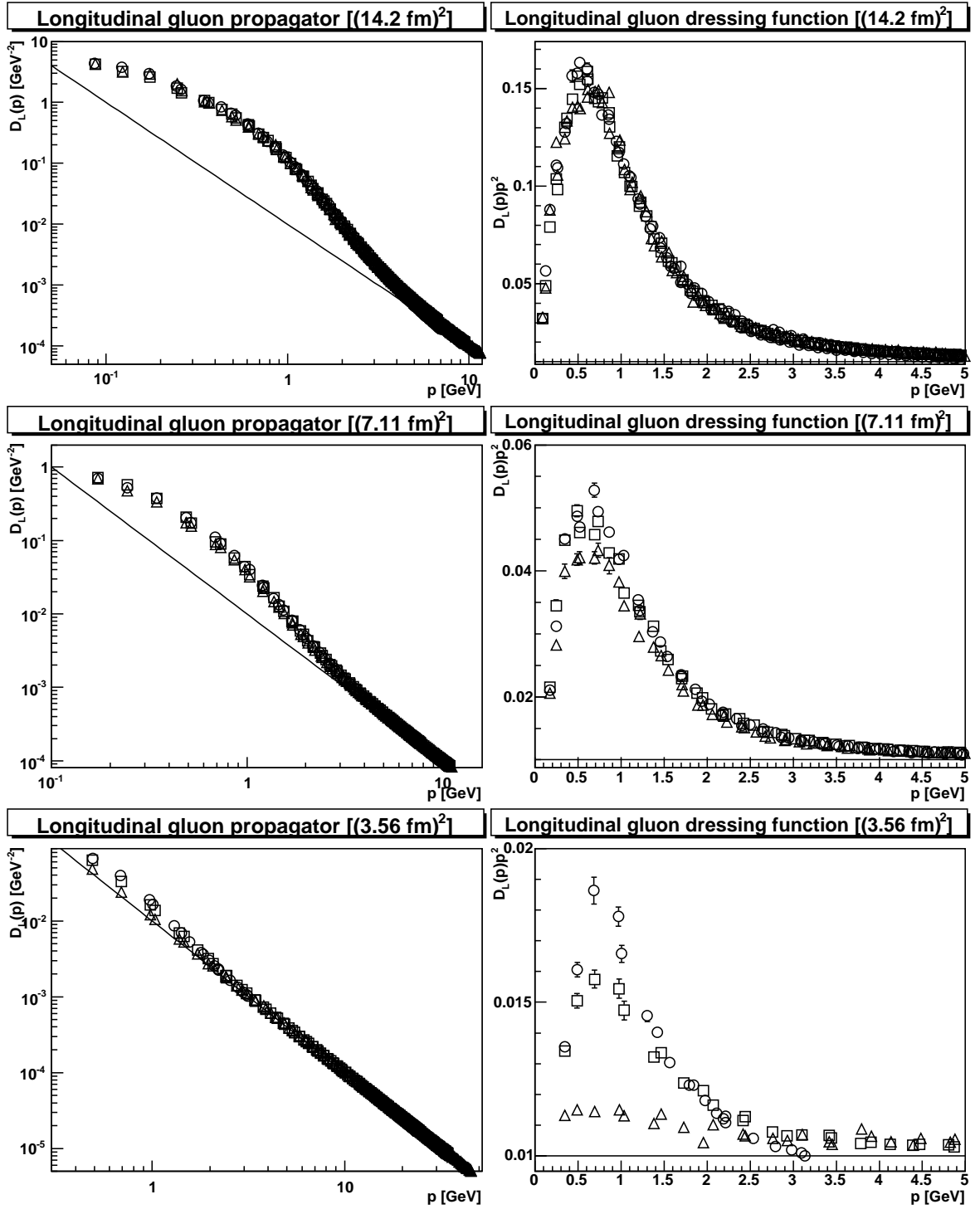}
\vskip -1mm
\caption{The longitudinal part of the gluon propagator for three different
lattice volumes $V$: (14.2 fm)$^2$ (top), (7.11 fm)$^2$ (middle) and (3.56 fm)$^2$ (bottom).
Symbols refer to different $\beta$ values in the three different panels, i.e.\
$\beta =$ 48.8 (Circles), 87.8 (Squares), 129.4 (Triangles) in the top panel; 
$\beta =$ 59.9 (Circles), 117 (Squares), 238.4 (Triangles) in the middle panel; 
$\beta =$ 10 (Circles), 1155 (Squares), 2505 (Triangles) in the bottom panel. All dressing
functions have been renormalized such that their lowest value coincides with the
renormalized continuum value $\xi=1/100$, indicated by a line. In all cases we considered
$n$=16 gauge copies for each
gauge orbit. Momenta are along the $x$-axis and along the $xt$-diagonal.}
\label{fgpl}
\end{figure*}

In the case of the two-point correlation function
\be
<\pd_\mu^x A_\mu^a(x) \, \pd_\nu^y A_\nu^b(y)>
\ee
the distribution should be proportional to $\delta^{ab}\delta_{\mu\nu}\delta(x-y)$. By applying
the Fourier-transform one finds the well-known Slavnov-Taylor identity (STI) that the
longitudinal part of the gluon propagator does not become dressed and is proportional to
$\delta^{ab}/p^2$, with the renormalized gauge parameter $\xi_r$ appearing in the constant
of proportionality \cite{Bohm:yx}
\be
p_\mu p_\nu D^{ab}_\mn(p)=\xi_r\delta^{ab}.\nn
\ee
Let us recall that this STI is a trivial consequence
of the gauge condition. Thus, its violation as a function of $\beta$ is a useful check
in order to control the extrapolation to the continuum limit.

Results for the longitudinal part of the gluon propagator, defined as\footnote{Here we
did not consider the mid-link Fourier transform of the gluon field \cite{Bowman:2002fe}.
However, the behaviors of the longitudinal gluon propagator for momenta along the $x$-axis
and along the $xt$-diagonal are rather similar (see Fig.\ \ref{fgpl}), suggesting that
discretization effects related to the definition of the gluon field in momentum
space are quite small.}
\be
D_L(p)=\frac{p_\mu p_\nu}{d(N_c^2-1)p^2}D^{aa}_\mn,
\ee
are reported in Fig.\ \ref{fgpl} for the case $\xi=1/100$. This value of $\xi$
was chosen, instead of the Feynman gauge $\xi=1$, in order to speed up the
approach to the continuum limit. One clearly sees that the dressing
function $p^2 D(p)$ is far from flat, but is becoming slowly flatter with increasing $\beta$.
Only at large momenta does one find an essentially constant function.
However, also at large momenta, renormalization effects are clearly visible. They can be
quantified by evaluating the renormalization factors $Z_\xi$ using the relation and renormalization condition
\be
\xi_r=Z_\xi \, \xi_m=\xi \, .
\label{eq:zxi}
\ee
Results are reported in Table \ref{tlzxi}. The increase of $Z_\xi$ with $\beta$ is consistent with
a logarithmic evolution, in accordance with the perturbative expectation. Moreover, this
evolution depends on the considered physical lattice volume, being faster for larger volumes.
As $Z_\xi$ tends to infinity in the limit $a\to 0$, the attraction of the Landau gauge as
a fixed point is evident.

\begin{table}
\begin{center}
\begin{tabular}{|c||c|c|c||c|c|c||c|c|c|}
\hline
$N$  & $\beta$ & $L$ [fm] & $Z_\xi$ & $\beta$ & $L$ [fm] & $Z_\xi$ & $\beta$ & $L$ [fm] & $Z_\xi$ \cr
\hline
20   & 10      & 3.55     & 1.003(2)&         &          &         &         &          &         \cr
\hline
30   & 21.95   & 3.55     & 1.010(2)&         &          &         &         &          &         \cr
\hline
40   & 38.6    & 3.55     & 1.013(2)& 10      & 7.11     & 1.030(7)&         &          &         \cr
\hline
50   & 60      & 3.55     & 1.013(1)& 15.35   & 7.11     & 1.00(1) &         &          &         \cr
\hline
60   & 86.3    & 3.55     & 1.014(3)& 21.9    & 7.11     & 1.031(3)&         &          &         \cr
\hline
70   & 117.4   & 3.55     & 1.012(2)& 29.63   & 7.11     & 1.045(3)&         &          &         \cr
\hline
80   & 153     & 3.55     & 1.013(1)& 38.55   & 7.11     & 1.053(2)& 10      & 14.2     & 1.017(8)\cr
\hline
90   & 193.7   & 3.55     & 1.011(3)& 48.65   & 7.11     & 1.061(3)& 12.55   & 14.2     & 1.02(1) \cr
\hline
100  & 239     & 3.55     & 1.011(1)& 59.9    & 7.11     & 1.064(1)& 15.4    & 14.2     & 1.088(8)\cr
\hline
120  & 344     & 3.55     & 1.011(1)& 86.1    & 7.11     & 1.062(3)& 21.9    & 14.2     & 1.203(4)\cr
\hline
140  & 468     & 3.55     & 1.014(1)& 117     & 7.11     & 1.064(1)& 29.7    & 14.2     & 1.231(5)\cr
\hline
160  & 611     & 3.55     & 1.018(2)& 152.7   & 7.11     & 1.067(2)& 38.6    & 14.2     & 1.248(3)\cr 
\hline
180  & 773     & 3.55     & 1.018(2)& 193     & 7.11     & 1.073(3)& 48.8    & 14.2     & 1.21(1) \cr
\hline
200  & 954     & 3.55     & 1.024(3)& 238.4   & 7.11     & 1.072(3)& 60.1    & 14.2     & 1.272(8) \cr
\hline
220  & 1155    & 3.55     & 1.022(3)& 288     & 7.11     & 1.079(4)& 72.6    & 14.2     & 1.277(6)\cr
\hline
242  & 1397    & 3.55     & 1.026(4) & 349     & 7.11     & 1.083(9)& 87.8    & 14.2     & 1.294(5)\cr
\hline
266  & 1689    & 3.55     & 1.037(7)& 421     & 7.11     & 1.078(6)& 103     & 14.2     & 1.306(6)\cr
\hline
294  & 2062    & 3.55     & 1.038(8) & 514.6   & 7.11     & 1.093(9)& 129.4   & 14.2     & 1.31(1) \cr
\hline
324  & 2505    & 3.55     & 1.05(1) & 625     & 7.11     & 1.12(1) & 157     & 14.2     & 1.31(3) \cr
\hline
\end{tabular}
\end{center}
\vskip 3mm
\caption{The renormalization constant $Z_\xi$ defined in Eq.\ \pref{eq:zxi},
obtained by requiring the lowest value of the longitudinal dressing function to
coincide with $\xi = 1/100$. Here we did not check for possible $Z(2)$ effects,
which strongly affect the Landau gauge gluon propagator for large $\beta$ values
\cite{Damm:1997vg}.}
\label{tlzxi}
\vskip 4mm
\end{table}

\begin{figure}
\includegraphics[width=\linewidth]{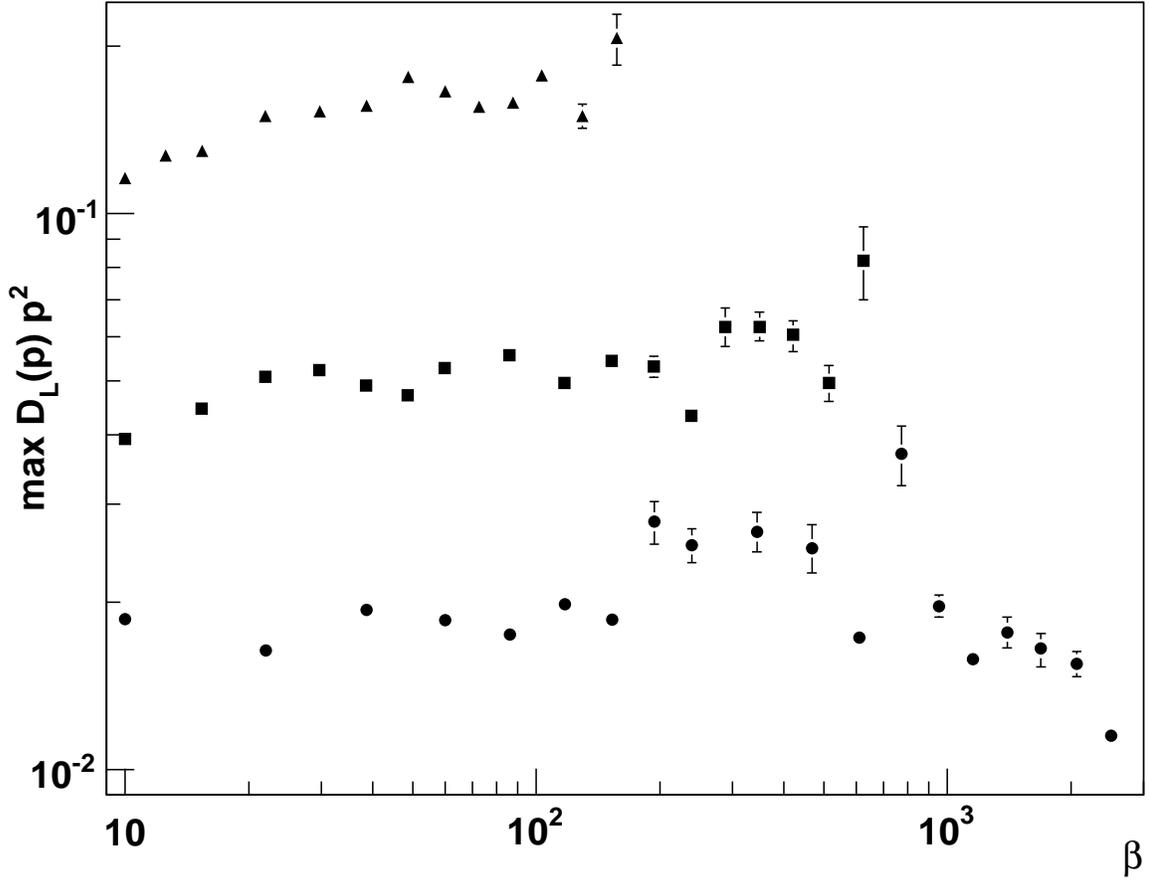}
\vskip 2mm
\caption{The $\beta$-dependence of the maximum of $p^2D_L(p)$ of the (renormalized) longitudinal
gluon propagator. Circles refer to the physical volume (3.55 fm)$^2$, squares to (7.11 fm)$^2$
and triangles to (14.2 fm)$^2$ (see Table \ref{tlzxi} for the corresponding $\beta$
values).}
\label{fbeta}
\vskip 4mm
\end{figure}

In any case, the main finite-$\beta$ effects can be seen at low momenta. Indeed, there is a clear
maximum in the longitudinal gluon dressing function for momenta of about 0.5--1 GeV
(see Fig.\ \ref{fgpl}).
We also find that this maximum seems to become flatter as $\beta$ increases, even though
this appears as a remarkably slow process. In addition, the height of the maximum increases
with the physical volume. The $\beta$ dependence of this maximum is shown in Fig.\ 
\ref{fbeta}. One sees that, for small $\beta$, the maximum seems to increase first,
before starting to decrease slowly to the expected continuum value.\footnote{This kind of
non-monotonic behavior has also been observed in the 2d Landau case when
considering finite-volume effects \cite{Maas:2007uv}.}
The slow evolution with $\beta$ is probably related to the fact that $\beta$ enters only
as a square-root in the width $\sigma$; it may also be influenced by the logarithmic
running of renormalization effects in two dimensions. The increase in strength of this
artifact as the volume increases is possibly due to the development of
would-be-zero modes of the Faddeev-Popov operator with volume. Projecting out
these modes from $\Lambda^a$ could reduce the strength of these discretization
artifacts.

It is also interesting to observe from Fig.\ \ref{fbeta} that the points characterized
by very large errors are also those that deviate most from the expected behavior.
The reason for this is a strongly asymmetric distribution of the
longitudinal part of the gluon dressing function as a function of configuration.
Particularly large values may be due to exceptional configurations \cite{Cucchieri:2006tf}.
At the same time, this could explain the large statistics required to obtain an essentially
continuous function of $\beta$ in Fig.\ \ref{fbeta}. Also, due to computational limitations,
we did not check if these results can be improved by increasing $n$ at large $\beta$.

Finally, we find similar qualitative results when considering a space-time dimension
$d$ larger than two.

\section{The transverse gluon propagator}\label{sgp}

\begin{figure*}
\includegraphics[width=\linewidth]{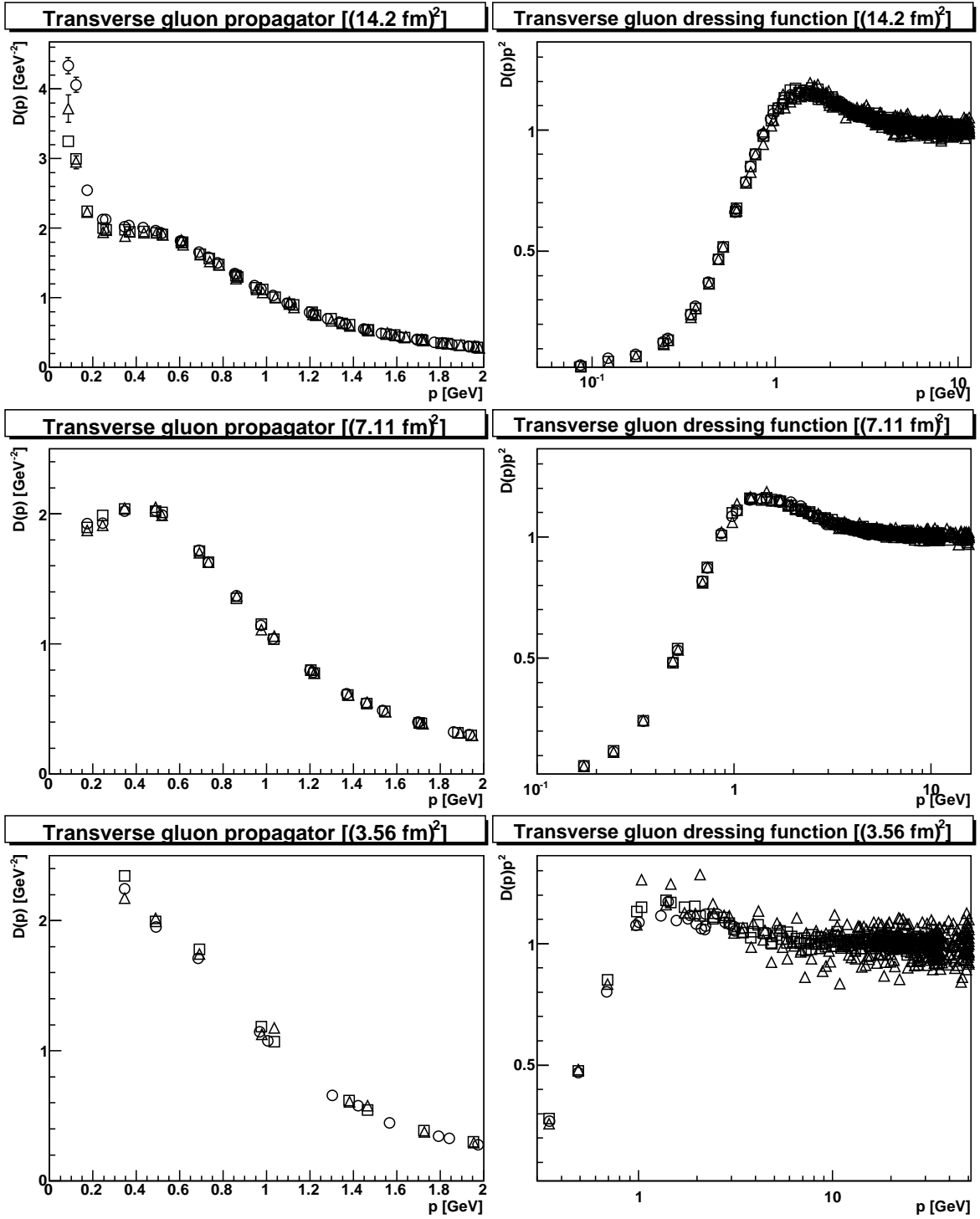}
\vskip 2mm
\caption{The transverse part of the gluon propagator for three different 
lattice volumes $V$: (14.2 fm)$^2$ (top), (7.11 fm)$^2$ (middle) and (3.56 fm)$^2$ (bottom).
Symbols refer to different $\beta$ values in the three different panels, i.e.\
$\beta =$ 48.8 (Circles), 87.8 (Squares), 129.4 (Triangles) in the top panel; 
$\beta =$ 59.9 (Circles), 117 (Squares), 238.4 (Triangles) in the middle panel; 
$\beta =$ 10 (Circles), 1155 (Squares), 2505 (Triangles) in the bottom panel.
Propagators and dressing functions have not been renormalized.
In all cases $\xi=1/100$ and we considered $n$=16 gauge copies for each
gauge orbit. Momenta are along the x-axis and along the xt-diagonal.}
\label{fgpt}
\vskip 4mm
\end{figure*}

As a first application of the gauge-fixing procedure introduced above, we consider
the numerical determination of the transverse part of the gluon propagator.
There are predictions \cite{Alkofer:2003jr,Sobreiro:2005vn} that,
at least in the 4d case, this propagator should be suppressed in the infrared
limit. 
Results in the 2d case for various volumes and $\beta$-values, using a standard
definition of the transverse gluon propagator \cite{Cucchieri:2006tf}, are shown in
Fig.\ \ref{fgpt} for $\xi = 1/100$, considering the same sets of configurations
used for the longitudinal gluon propagator reported in Fig.\ \ref{fgpl}.
We see that for momenta above about half a GeV there is no
pronounced volume or $\beta$-dependence, in contrast to the case of the longitudinal
propagator. In particular, the dressing function is qualitatively similar to the corresponding
dressing function in Landau gauge \cite{Maas:2007uv}. 
On the other hand, at small momenta (i.e.\
below half a GeV) we find finite-volume and $\beta$-effects. In particular, similarly
to the Landau-gauge case, there seems to be a maximum in the
gluon propagator for momenta of about 400 MeV, at least at large enough
physical volume. However, on the largest physical volume, there is also a rise
in the infrared (below about 250 MeV), which diminishes with increasing $\beta$. Thus,
the quantitative properties of this rise are strongly affected by discretization
effects, and a final conclusion on the infrared behavior cannot yet be drawn.

\section{Summary}\label{ssum}

We have described a new class of gauges on the lattice, which become the
linear covariant gauges in the formal continuum limit. The residual
${\cal O}(a)$ errors have been investigated in detail. We find that our gauge
approaches the correct continuum gauge, albeit slowly (especially at
large physical volumes). One should check if it is possible to speed up the approach
to the continuum by eliminating the leading discretization errors. This 
speed-up would probably be crucial in the study of the infrared sector of the
theory, in particular when considering space-time dimension $d$ larger than two.

As a first application we have presented a preliminary study of the transverse part of
the gluon propagator in two dimensions. Our results in the infrared limit are
not conclusive and more investigation is required in order to
obtain a clearer picture. Of course it would also be interesting to
extend this study to higher dimensions and to other quantities, such as the
ghost propagator. However, since the gauge field is no longer transverse, this
study is more involved than in the usual Landau gauge. Finally, one should also
check for possible Gribov-copy effects on the evaluated quantities.


{\bf Acknowledgments}
 
A.\ M. was supported by the DFG under grant number MA 3935/1-1 and MA 3935/1-2 and by the
FWF under grant number P20330.
A.\ C. and T.\ M. were partially supported by FAPESP (under grants \# 00/05047-5
and 05/59919-7) and by CNPq (including grant \# 476221/2006-4). The work of T.M. is
supported also by the Alexander von Humboldt Foundation.
The ROOT framework \cite{Brun:1997pa} has been used in this project.


\end{document}